\documentclass{article}

\usepackage[english]{babel}

\usepackage[letterpaper,top=2cm,bottom=2cm,left=3cm,right=3cm,marginparwidth=1.75cm]{geometry}

\usepackage{amsmath}
\usepackage{graphicx}
\usepackage{authblk}
\usepackage[colorlinks=true, allcolors=blue]{hyperref}
\usepackage{cleveref}

\usepackage{authblk}
\title{Can a Programmable Phase Plate serve as an Aberration Corrector in the Transmission Electron Microscope (TEM)?}
\author[1,2]{\small Francisco Vega Ibáñez}
\author[1,2]{\small Armand Béché}
\author[1,2,*]{\small Johan Verbeeck}
\affil[1]{\footnotesize University of Antwerp, EMAT, Groenenborgerlaan 171, 2020, Antwerp, Belgium}
\affil[2]{\footnotesize NANOlab Center of Excellence, University of Antwerp, Groenenborgerlaan 171, 2020 Antwerp, Belgium}
\affil[*]{\footnotesize Corresponding Author: Jo Verbeeck \href{mailto:jo.verbeeck@uantwerp.be}{jo.verbeeck@uantwerp.be}}
\date{}                     
\begin{document}
  \maketitle
 
\begin{abstract}
Current progress in programmable electrostatic phase plates raises questions about their usefulness for specific applications. Here, we explore different designs for such phase plates with the specific goal of correcting spherical aberration in the Transmission Electron Microscope (TEM). We numerically investigate whether a phase plate could provide down to 1 Ångström spatial resolution on a conventional uncorrected TEM. Different design aspects (fill-factor, pixel pattern, symmetry) were evaluated to understand their effect on the electron probe size and current density. Some proposed designs show a probe size ($d_{50}$) down to 0.66\AA~, proving that it should be possible to correct spherical aberration well past the 1\AA~ limit using a programmable phase plate consisting of an array of electrostatic phase shifting elements.
\noindent(Received 11 05 2022; revised 10 06 2022; accepted 28 06 2022)
\end{abstract}

\section{Introduction}
Aberration correction in the electron microscope is a topic that started when, in 1936, Otto Scherzer proved that chromatic and spherical aberrations are unavoidable in cylindrically symmetric static electron lenses (later on called "Scherzer Theorem" due to its importance) \cite{Scherzer1936}. Shortly after, in 1947, Scherzer realized that using an array of multipolar lenses could allow phase manipulation beyond his theorem's rotationally symmetric constraints, thus providing a pathway for correcting the formerly mentioned aberrations \cite{Scherzer1947}. Attempts toward experimental realization saw many iterations throughout the following decades \cite{seeliger1951spharische,archard1954requirements,mollenstedt1954elektronenmikroskopische,meyer1961praktische,kelman1962achromatic,dymnikov1963quadrupole,hardy1969combined,rose1971elektronenoptische,koops1977test,bernhard1980erprobung}. However, the successful implementation of an aberration corrector for TEM did not come until 50 years later with the Haider-Rose-Urban project \cite{Haider1995CorrectionOT,Haider1998ElectronMI,Rose1990Outline}. The implementation of this corrector represented a breakthrough in the field, allowing for beyond-Ångström resolution in both TEM and STEM and up to atomic resolution in analytical methods due to a significantly increased current density.\\
More than 20 years after this breakthrough in the field of TEM, we want to explore the possibility of implementing a different idea to correct for third-order spherical aberration ($C_{s}$). To achieve this, we take inspiration from the field of optics, more specifically spatial light modulators, that allow to freely program the wavefront of coherent light, making use of a range of different technologies \cite{efron1994spatial,maurer2011spatial}. "Spatial Electron Modulators," as opposed to their light optic counterparts, are unfortunately still far from achieving a similar level of technological advancement.\\
The concepts and reasons why a programmable phase plate is attractive in general have been discussed earlier \cite{Guzzinati_2015,de2021optical}. We have opted for a technological approach that uses an array of electrostatic Einzel lenses as programmable phase shifters. We demonstrated a 2x2 proof of concept \cite{verbeeck2018demonstration} and have since then put significant efforts into extending the concept to the current state of the art where a 48-pixel programmable phase is produced by lithographic means \cite{Verbeeck2021}. In the meanwhile, other groups have explored different means to achieve similar freedom in phase shaping of electron beams using miniaturized multipolar lenses \cite{grillo2014generation}, interaction with optical near fields \cite{de2021optical, wang2020coherent,maxson2015adaptive,constantin2022transverse,konevcna2020electron}, electrostatic nanofabricated elements applying a projected potential to a region of free space \cite{schultheiss2006fabrication,nagayama2009phase,hettler2012improving,cambie2007design,beche2017efficient}, and many others.\\
This technological progress allows to ponder potential applications such phase plate could provide. In this paper, we will focus on whether an array of electrostatic phase shifters could be made into an acceptable $C_{s}$ corrector for TEM. Proving this to be possible, would provide the option of introducing a small, integrated device into the TEM column with little change to its configuration. Furthermore, it would offer a rapid response tool, which combined with adaptive algorithms can auto-tune and respond to instrument or specimen-induced drifts, potentially reducing the overall complication of an experiment (dose and material efficiency).\\
As simple as the idea may sound, the devil is in the details, and we attempt in this paper to give an overview of the design parameters that have to be balanced between manufacturability and expected performance to evaluate if aberration correction with a programmable electrostatic phase plate could have a future in TEM.

\section{Methods}
\subsection{Electron Beam Parameters to be Optimized}

\begin{figure}
\centering
\includegraphics[width=\textwidth]{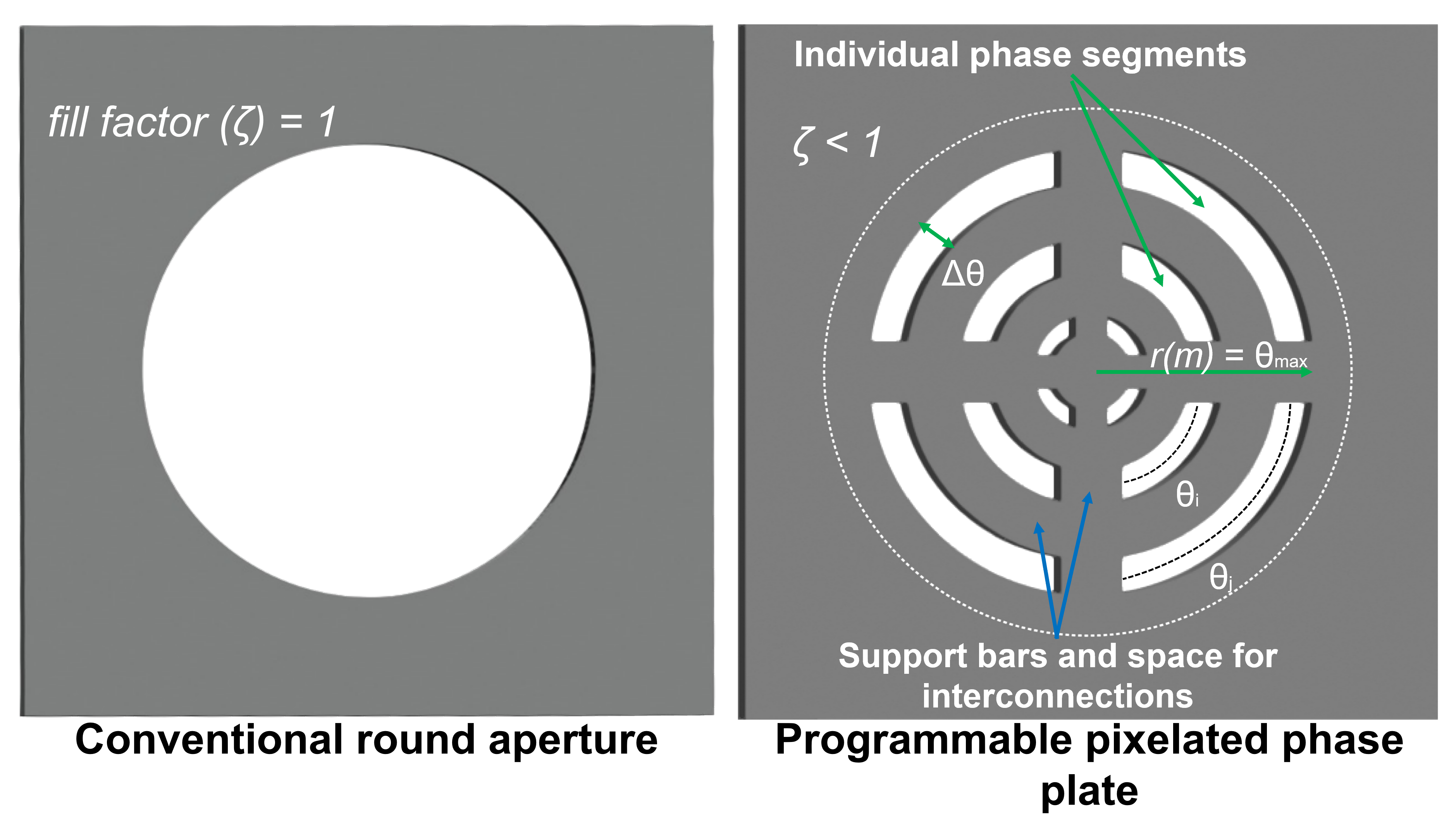}
\caption{\label{fig:phaseplatesketch}Sketch of a conventional round aperture versus an array of phase-shifting elements occupying a similar area as in the round aperture case. Parameters such as fill factor ($\zeta$) and angular range ($\Delta\theta$) are indicated along with the relation between maximum aperture ($\theta_{max}$) and radius ($r(m)$). We show where the interconnections and supports are allocated in the aperture, thus reducing the fill factor.}
\end{figure}

In order to evaluate different designs for programmable phase plates, we need to first agree on the beam parameters to optimize.\\
For aberration correction, we are interested in spatial resolution and current density in the electron probe. It is convenient to define the spatial resolution as $d_{50}$, the diameter of the probe containing 50\% of the beam intensity \cite{kohl1985theory}. This definition comes very close to the FWHM for very sharp beams, while it offers the advantage of accounting for the effect of the beam tails. For the case of current density, we assume the phase plate to be homogeneously and coherently illuminated with a current density that would lead to a total probe current of $I_0$ if a circular aperture would replace the phase plate with the same total diameter. As the electrostatic phase plate will block part of this beam inherent to the construction of the segments making up for it (sketched in \cref{fig:phaseplatesketch}), we get for the beam current with phase plate $I'=I_0  \zeta$ with $\zeta$ the \emph{fill factor} of the specific phase plate. Ideally, this fill factor should be as close to 1 as possible, meaning no blocking of the electron beam, but practical design constraints will determine what is realistic to achieve.\\
The average current density, $J_{50}$, in the $d_{50}$ probe diameter is then given as:
\begin{eqnarray}
J_{50}= \frac{4 \zeta I_0 }{\pi d_{50}^2} 
\label{currdens}
\end{eqnarray}
In comparison, for an ideal aberration-corrected system with circular aperture and convergence half angle $\alpha$, this becomes:
\begin{eqnarray}
d_{50,ideal}&\approx&0.514\frac{\lambda}{ \alpha}\\
J_{50,ideal}&\approx&15.152\frac{I_0 \alpha^2}{\pi \lambda^2} 
\end{eqnarray}
Note that in this theoretical design exercise, we ignore other sources of experimental probe broadening, such as those caused by source size \cite{michael1987consistent}, vibrations, partial coherence, and any sources of electronic noise or thermal drifts that could affect the final probe size \cite{von1995instrumental}.\\ 
To keep the arguments as general as possible, we focus on high-level design parameters and avoid going into details and technical challenges arising from manufacturing.

\subsection{Phase Plate Design Parameters}
After listing the probe parameters we aim to study and optimize, we now will give an insight into some phase plate design considerations to achieve probe size reduction and current density increase. We want to pay special attention to the number of phase-shifting elements, their width, and the percentage of the aperture they will block (all correlated through the interconnections that deliver the potential to each individual element). A simple rule of thumb would be that more segments come with more interconnections, thus blocking more of the incoming electron beam. However, as mentioned before, we will discuss the mathematical implications of the segment/phase plate design leaving the manufacturing of the phase plate itself out of the scope of this study.

\subsubsection{The Role of the Fill Factor}

Putting material in the path of the beam is inevitable for these phase plates since it is the only way to break the symmetry constraints imposed by the Maxwell equations on fields in free space. Adding material to create the structure of the plate segments leads to the concept of the fill factor ($\zeta$), the amount of optically transparent versus non transparent parts in the phase plate. This modulation of the local amplitude of the electron beam will inevitably lead to a broadening of the probe as high spatial frequency tails are introduced.\\
If we try to estimate this effect, we can begin with a wave function on the probe forming aperture with a constant amplitude:
\begin{eqnarray}
\psi_{in}(\vec{k})=\frac{1}{\sqrt{A}} a(\vec{k}) e^{i \phi(\vec{k})}
\end{eqnarray}
With $a(\vec{k})$ a function defining the aperture's shape being either 1 ($\vec{k}<\alpha$) or 0 ($\vec{k}>\alpha$), $A=\int a(\vec{k}) d^2\vec{k}$ is the total area of the aperture, $\phi$ the local phase, and $\vec{k}$ a vector in the aperture plane.
The wave function in real space then becomes:
\begin{eqnarray}
\Psi\left(\vec{r}\right)=\int \psi_{in}(\vec{k})e^{i \vec{k}\cdot \vec{r}} d^2 \vec{k}
\end{eqnarray}
We consider as the ideal case a circular aperture and a flat phase (i.e., the diffraction limit). In this case, the probe will be the sharpest and the wave will have a value at the central point:
\begin{eqnarray}
\Psi_{ideal}\left(0\right)=\frac{1}{\sqrt{A}}\int a(\vec{k})d^2 \vec{k}=\sqrt{A}
\end{eqnarray}
Now, if we want to describe the situation of a pixelated phase plate with the same outer dimensions, we can put a mask $M$ over the ideal aperture, which is either 1 (electron transparent) or 0 (not electron transparent). If we assume the ideal case where the pixelated phase plate can still provide a flat phase in those areas where the plate is electron transparent, we get:
\begin{eqnarray}
\psi_{pp}(\vec{k})=\psi_{in}(\vec{k}) M(\vec{k})
\end{eqnarray}
This mask changes the maximum of the real space wave function to:
\begin{eqnarray}
\Psi_{pp}(0)=\sqrt{A} \zeta
\end{eqnarray}
With $\zeta=\int \frac{M(\vec{k})}{a(\vec{k})}  d^2\vec{k}$ the fill factor of the phase plate.\\
Now, let us consider the resulting probe which consists of the sum of the ideal corrected wave and an unwanted tail part:
\begin{eqnarray}
\Psi_{pp}(\vec{r})=\zeta \Psi_{ideal}(\vec{r})+\Psi_{tails}(\vec{r})
\end{eqnarray}
Where the scale factor $\zeta$ describes the scaling of the central maxima with respect to the ideal corrected case.
We can now write the intensity of the probe as:
\begin{eqnarray}
I_{pp}(\vec{r})\approx \zeta^{2} I_{ideal}(\vec{r})+I_{tails}(\vec{r})
\end{eqnarray}
We assume that the current of the tails does not overlap with the central spot. This assumption is reasonable given that the ideal probe is a maximally compact function near the center, and the tails come from the high spatial frequencies of the mask, which are much smaller than the total aperture radius.\\
If we normalize the total current illuminating the round aperture $I_{ideal,total} = I_0 =1$ for simplicity, the total intensity in the probe then becomes:
\begin{eqnarray}
\begin{aligned}
I_{tot,pp} &\approx \zeta^2 + I_{tot,tails}\\
\zeta &\approx \zeta^2 + I_{tot,tails}\\
I_{tot,tails} &\approx \zeta(1-\zeta)
\end{aligned}
\end{eqnarray}
If we normalize the tails relative to the total intensity in the probe, we get:
\begin{eqnarray}
I_{tot,tails,rel} \approx 1-\zeta
\end{eqnarray}
In other words, the unwanted tail part of the probe formed by a pixelated phase plate scales approximately as $1-\zeta$. These tails will form a low-resolution background signal to any scanned probe setup. This background is highly unwanted as it will increase the counting noise, which is especially bad for spectroscopic methods since it will bring signal from areas away from the probe center. To prevent these tails, we want to create a mask having the highest possible fill factor. For the same reason, to optimize the value of $d_{50}$, we need a $\zeta>0.5$, and the ideal case would be to bring this value as close to 1 as possible. Cutting off the tails with an aperture placed lower in the TEM column seems another option, but this would require cutting apertures with an equivalent real space diameter only a few orders larger than the probe size, which seems extremely difficult to obtain if we aim for \AA~ probes, especially when considering that working in another (magnified) plane than the sample plane will introduce inevitable lens aberrations.

\subsection{Phase Plate Pixel Pattern}
In order to best compensate for the lens aberrations in a pixelated phase plate, it is important that each phase-changing segment can locally correct for the phase error of the other lenses in the microscope as well as possible. This will naturally lead to pixel patterns that will mimic the symmetry of the aberration function. Starting from the aberration function $\chi(\theta)$ and considering only the defocus ($\Delta f$) and $C_{s}$ terms we have:
\begin{eqnarray}
\chi(\theta)=\frac{\pi}{\lambda} \left[- \Delta f \theta^2 + \frac{C_s}{2} \theta^4\right]\end{eqnarray}
We now look for the highest angle that still can be corrected by a segment in the phase plate, and we assume a spherically symmetric phase plate segment covering an angular range between $\theta_i$ and $\theta_{i+1}$. 
Taylor expanding the aberration function to second-order around $\theta_i$ gives:
\begin{equation}
\begin{aligned}
\chi|_{\theta_i}(\Delta \theta) \approx& \frac{\pi}{\lambda} \left[- \Delta f \theta_i^2 + \frac{C_s}{2} \theta_i^4\right] + \\
&\frac{2\pi}{\lambda} \left[- \Delta f \theta_i + C_s \theta_i^3\right]\Delta\theta + \\
&\frac{\pi}{\lambda}\left[- \Delta f + 3C_s \theta_i^2\right]{\Delta\theta}^2+ \ldots
\end{aligned}
\end{equation}
\subsubsection{Zeroth-Order Phase Correction}
Suppose we use a zeroth-order phase plate which produces a constant phase shift that is programmable per segment. If we allow for a maximum phase error $\epsilon$ within each segment, we get for the maximum angle up to which we can correct:
\begin{equation}
\chi|_{\theta_i}(\Delta \theta)<\epsilon
\end{equation}
\begin{equation}
\frac{\pi}{\lambda}\bigl[\left(-2\Delta f\theta_i+2C_s\theta_i^3\right)\Delta\theta+
\left(-\Delta f+3C_s\theta_i^2\right){\Delta\theta}^2\bigr]<\epsilon
\end{equation}
This puts an upper limit on the maximum angle that can be corrected depending on how small we can make $\Delta \theta$.
If we assume only $C_s$ needs correction, we can always choose $\Delta f=0$, and we also take the first-order Taylor expansion as sufficient we get:
\begin{eqnarray}
\label{delta_zero}
\Delta\theta < \frac{\epsilon \lambda}{\pi 2C_s \theta_i^3 }
\end{eqnarray}
For a typical $C_s=1$~mm, $\theta_i$=15~mrad,  and $\epsilon={2\pi}/{10}$, we obtain a $\Delta\theta < 58\mu rad$. This would require feature sizes of the segments of only 0.3\% of the total aperture diameter and could become rather difficult to manufacture.
Alternatively, we can express the maximum angle for a given minimum size of $\Delta \theta$:
\begin{eqnarray}
\theta_{max}< \left(\frac{\epsilon \lambda}{2 \pi C_s \Delta\theta }\right)^\frac{1}{3}
\end{eqnarray}
This leads to 5.8~mrad for $\Delta \theta=1$~mrad, giving us the maximum aperture angle we can correct with a flat phase within the given error $\epsilon$.

\subsubsection{First-Order Phase Correction}
\begin{figure}
\centering
\includegraphics[width=\textwidth]{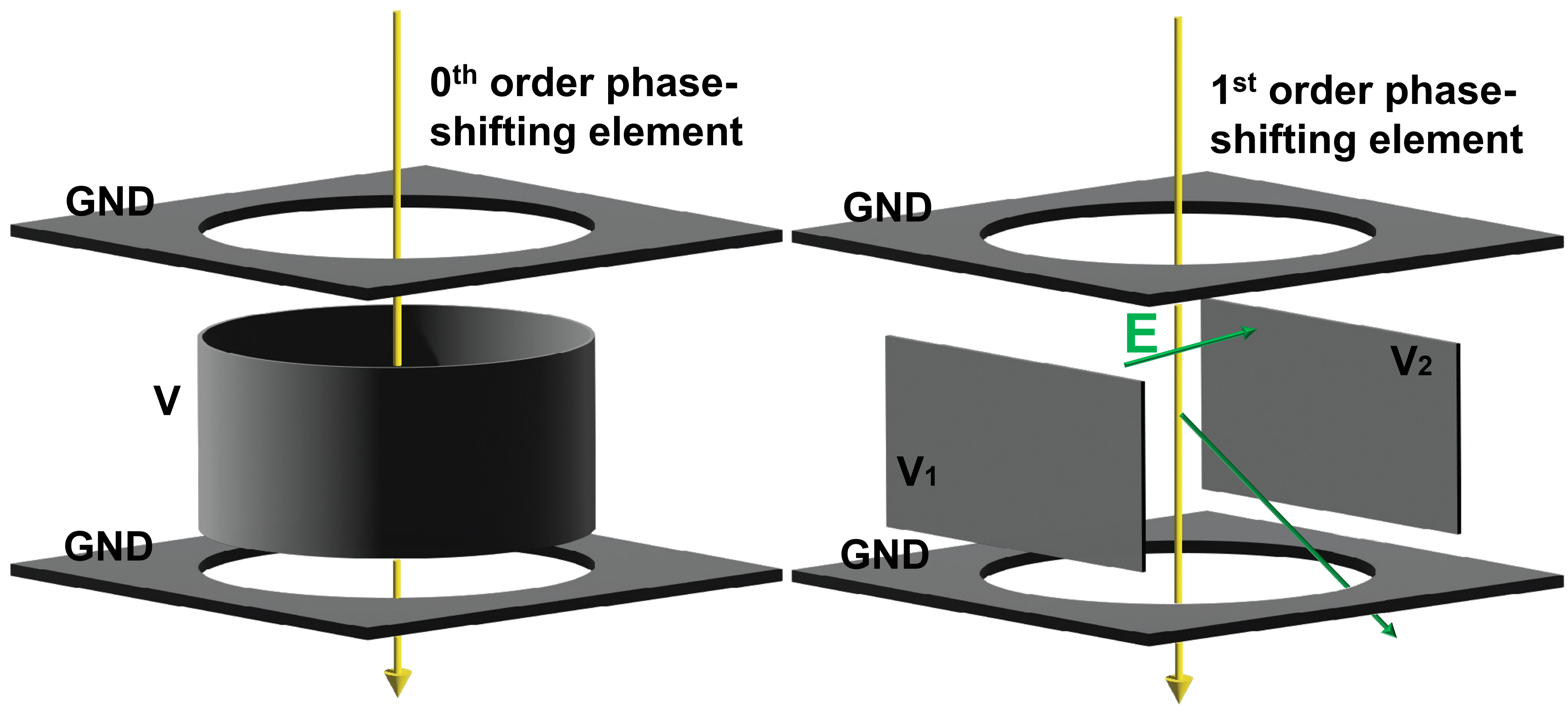}
\caption{\label{fig:0 vs 1 st order}Sketch of zeroth- and first-order phase element as the main building blocks of an array of programmable phase-shifting segments.}
\end{figure}

\begin{figure}
\centering
\includegraphics[width=\textwidth]{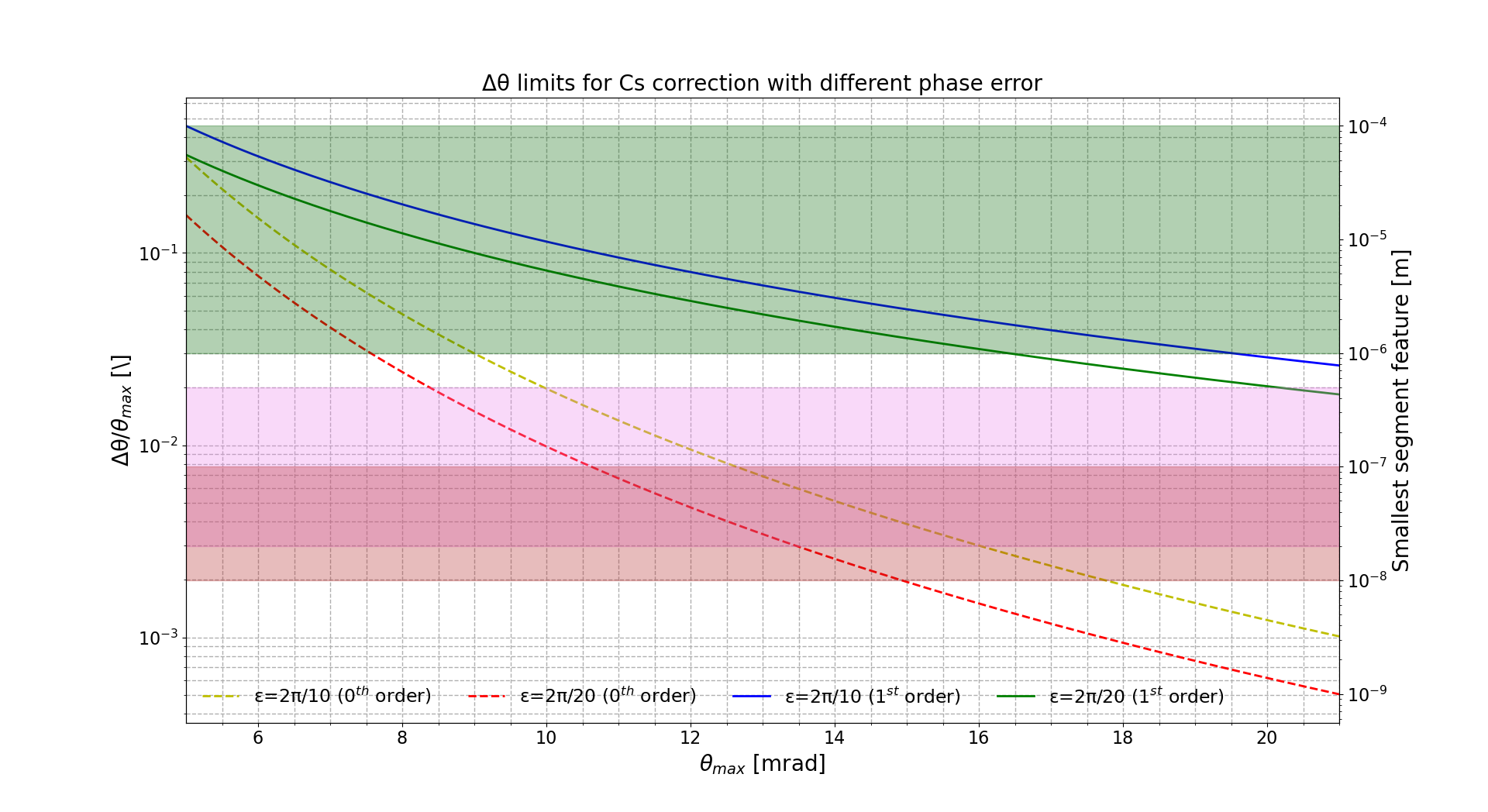}
\caption{\label{fig:min. feature size} Minimum $\Delta\theta$ needed to correct for $C_{s}$ as a percentage of the total aperture for different phase errors $\epsilon$ (left axis, log scale). These can be translated to minimum segment feature sizes when assuming a total aperture diameter of 70~$\mu$m (right axis). The shaded regions (green, violet, red) show approximated region where photolithography, extreme ultra violet (EUV) lithography, and electron beam lithography (EBL) would be required to make such features \cite{engstrom2014additive}.}
\end{figure}

If, on the other hand, we allow for first-order correction in each phase segment, meaning a linear projected potential ramp in the radial direction and thus requiring at least two independent potential electrodes per segment, the situation changes. In this case the phase could be corrected up to first order and we get as phase error:
\begin{eqnarray}
&\chi|_{\theta_i}(\Delta \theta) <\epsilon\\
&\frac{\pi}{\lambda}|\left(-\Delta f + 3 C_s \theta_i^2 \right)|{\Delta\theta}^2 < \epsilon \\
&\Delta\theta < \sqrt{\frac{\epsilon \lambda}{\pi\left(\Delta f + C_s 3\theta_i^2\right)}}
\end{eqnarray}
We choose $\Delta f=0$ , which yields:
\begin{eqnarray}
\label{delta_first}
\Delta\theta < \sqrt{\frac{\epsilon \lambda}{3 \pi C_s \theta_i^2}}
\end{eqnarray}
For a typical $C_s=1$~mm, $\theta_i$=15~mrad, and $\epsilon={2\pi}/{10}$ we get $\Delta\theta < 0.76$~mrad which is $\approx$13 times larger as compared with zeroth-order correction.
Following the steps of the previous section, we can express the maximum angle for a given minimum size of $\Delta \theta$:
\begin{eqnarray}
\theta_{max}< \sqrt{\frac{\epsilon \lambda}{3 \pi C_s \Delta\theta ^2}}
\end{eqnarray}
This leads to 11.46~mrad for $\Delta \theta=1$~mrad, nearly double its zeroth-order counterpart.\\
We give a simplified sketch of the main building blocks needed to make up for both a zeroth- and first-order phase-shifting elements in \cref{fig:0 vs 1 st order}. Furthermore, we show a plot of eq. \ref{delta_zero} \& eq. \ref{delta_first} in \cref{fig:min. feature size} for two different phase errors $\epsilon$. To put this into perspective, we give the resolution ranges for some manufacturing techniques (shaded regions).\\
In order to translate the previous results to meters, we can take a scaling factor to relate angle (mrad) and physical distance (meters), assuming that the widest area we can coherently illuminate is in the order of 100$\mu$m (so, despite the maximum aperture angle $\theta$, we still illuminate the same area in meters). With this in mind, and looking at the right axis scale on \cref{fig:min. feature size}, we can get a value for the physical dimension corresponding to the minimum $\Delta\theta$ needed to keep the phase error under a specific error ($\epsilon$).

\section{Results}
\begin{figure}
\centering
\includegraphics[width=1\textwidth]{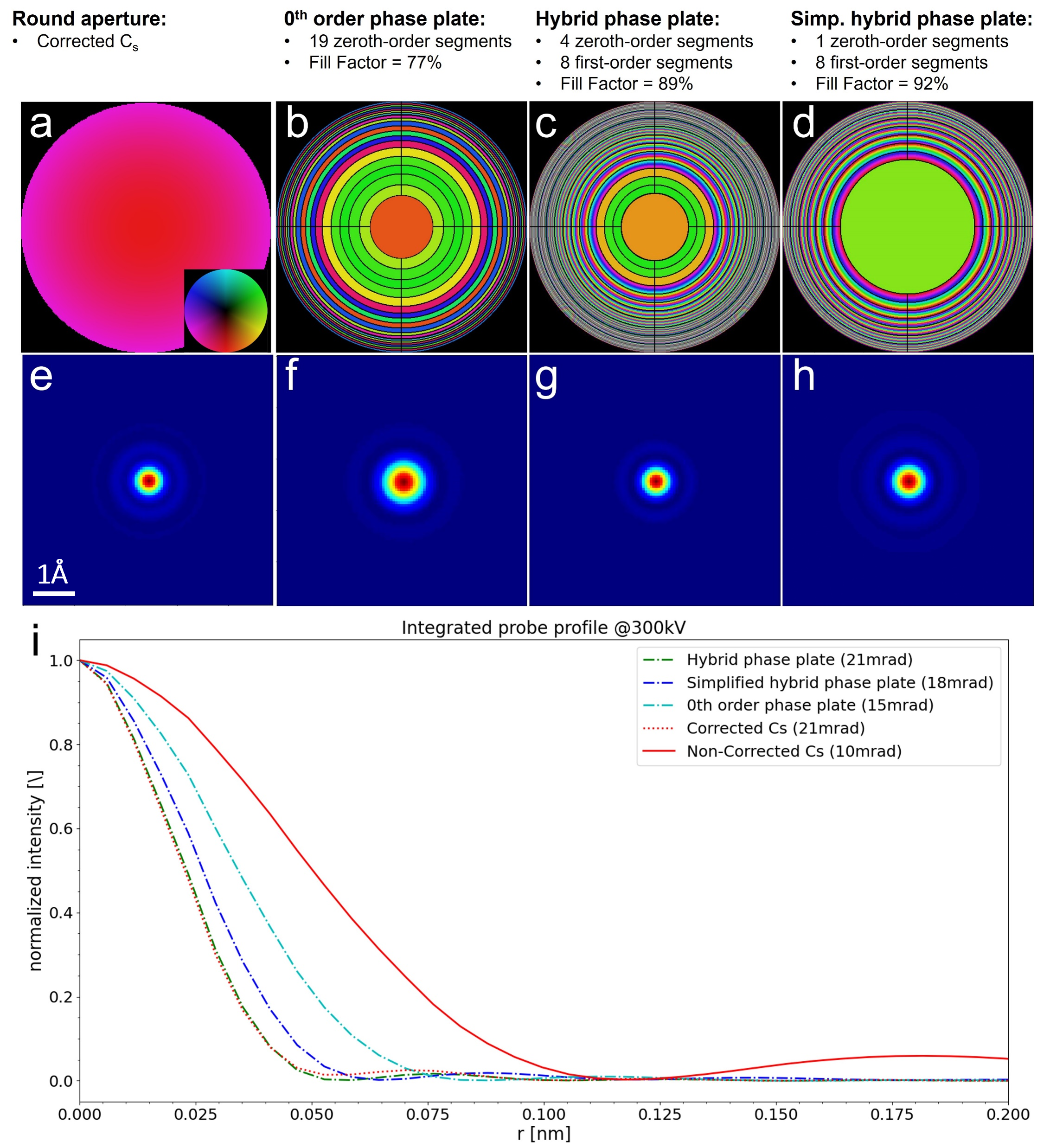}
\caption{\label{fig:probe profile}Simulated performance of different apertures at 300~keV. (a) The aberrations over a corrected round aperture at 21~mrad with $ C_{s} = 1\mu m$ and Scherzer's defocus. The aberrations for (b,c,d) are 1.2~mm $C_{s}$ with Scherzer's defocus as well. (b) A zeroth-order phase plate with 19 segments and $\approx 77\%$ fill factor at 15~mrad opening angle, (c) and (d) hybrid correction phase plates with four central zeroth-order segments followed by eight first-order segments and $\approx89\%$ fill factor at 21~mrad in the case of (c), and one central zeroth-order hole followed by eight first-order segments and $\approx92\%$ fill factor at 18~mrad for (d). (e-h) The simulated probe intensities below the corresponding phase plate responsible for them; the simulation box is  6x6\AA~. (i) A radially integrated profile for each of the abovementioned figures. The proposed alternatives improve the spot size compared with the aberrated instrument. However, it is important to mention that the feature size of the smallest segment in (b) is $\approx 270nm$, whereas the hybrid plates' segments are $1.5\mu m$ wide. The color wheel inset in (a) shows the scale used to represent both phase (hue) and amplitude (intensity).}
\end{figure}

\begin{figure}
\centering
\includegraphics[width=\textwidth]{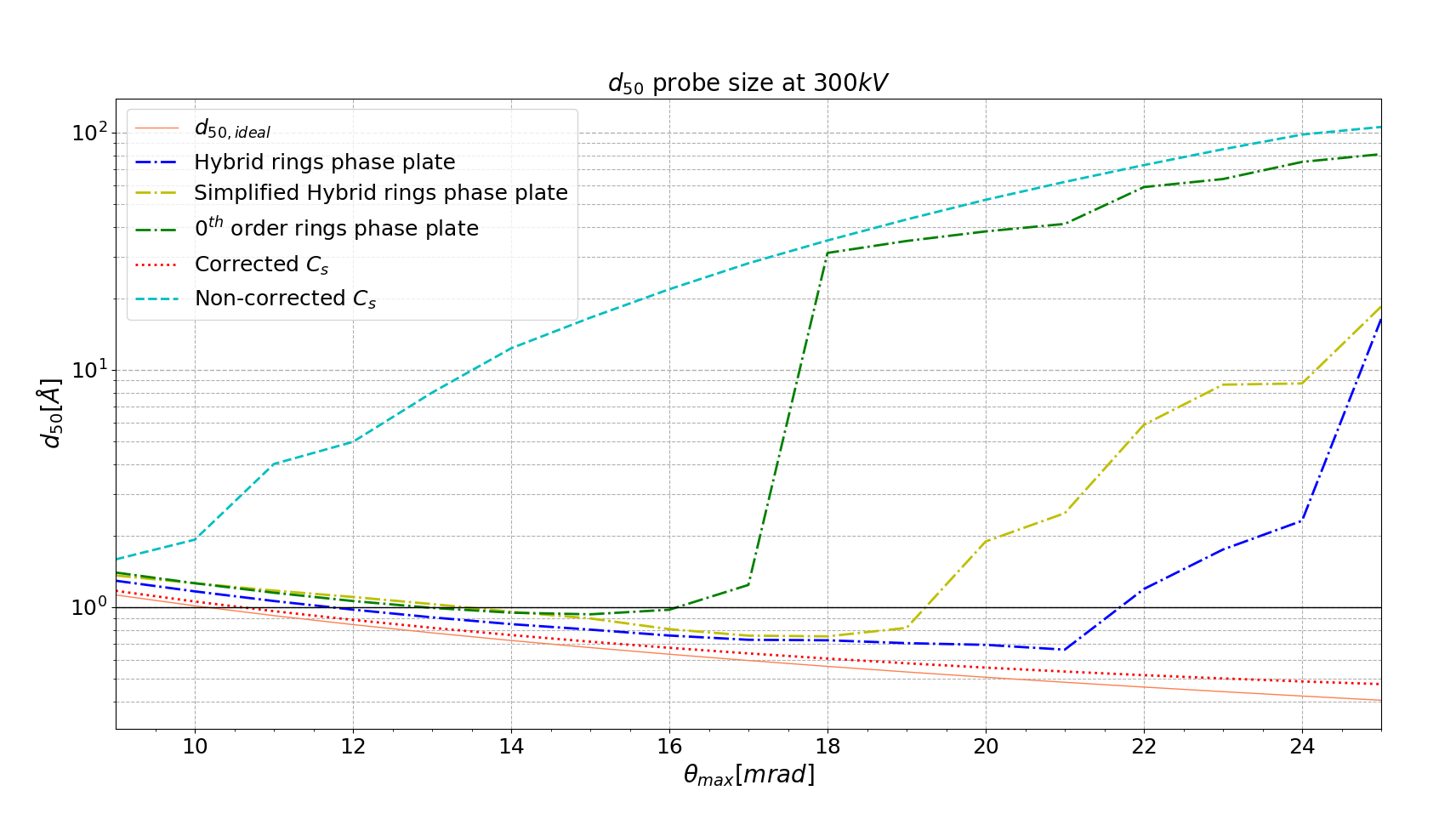}
\caption{\label{fig:probe size}Simulated probe size $d_{50}$ assuming 300~keV, $C_s$ = $1.2$~mm, and Scherzer defocus for the phase plates and non corrected aperture. The black line shows the 1\AA~ limit. All proposed phase plate designs are capable of a probe size below this limit (blue, yellow, green lines). However, they are still outperformed by a multipole corrector (red line).}
\end{figure}

\begin{figure}
\centering
\includegraphics[width=\textwidth]{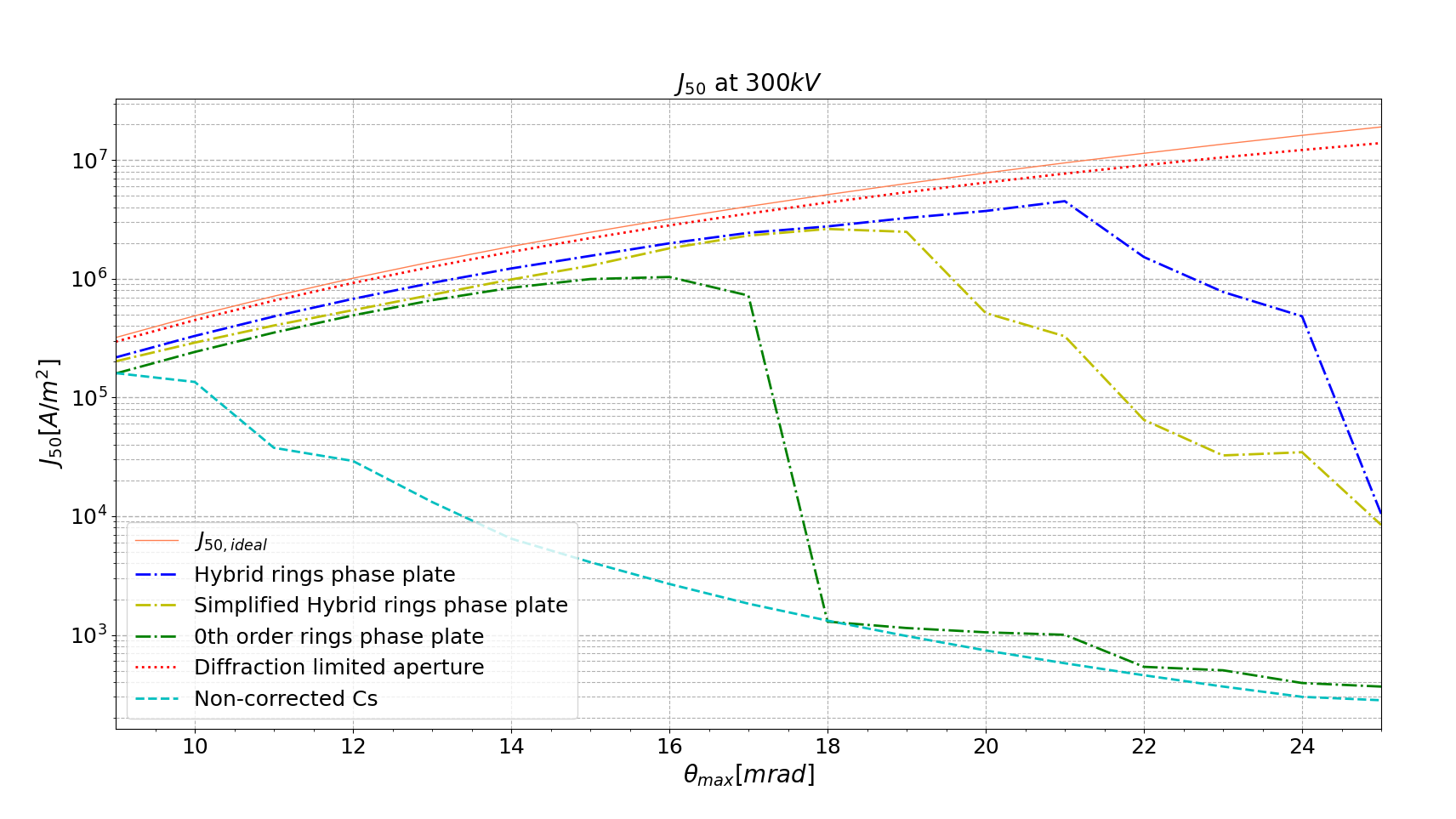}
\caption{\label{fig:current density}Current density $J_{50}$ for the different phase plate designs at different $\theta_{max}$ assuming a total incoming current up to 50~pA. All proposed designs significantly improve compared with a non corrected aperture; the red (dotted) line shows the corrected instrument's performance.}
\end{figure}

Integrating the design rules discussed above, we numerically simulated a set of different electrostatic phase plate designs to test their capabilities to correct $C_s$ at 300~keV with two approaches, (1) applying constant phase shift (zeroth-order) and (2) a combination of constant and linear ramp shift (first-order) segments. As $C_{s}$ is rotationally symmetric, the proposed designs all consist of concentric rings to make maximum use of the symmetry of the problem.\\
For reference, these proposed concentric segments shown in \cref{fig:probe profile} (a-d) are analogous to those labeled in \cref{fig:phaseplatesketch} as "individual phase segments," with the only difference being that we reduce the spacing between segments arbitrarily for our study, and their working principle is the same as the one sketched in \cref{fig:0 vs 1 st order}.\\
After some design iterations, we narrowed down the study to compare three apertures: zeroth-order concentric rings (\cref{fig:probe profile} b), a hybrid design (\cref{fig:probe profile} c), and a simplified version of the latter (\cref{fig:probe profile} d). We show the resulting probe profiles from the apertures in \cref{fig:probe profile} (i). This figure shows how the probe from all proposed designs approaches that of a corrected instrument, visibly improving over a non corrected instrument. It is essential to mention that \cref{fig:probe profile} (i) only gives a view of an azimuthally integrated intensity which is normalized to the maximum intensity of each probe for scale (y-axis) and shows only the tails of the lower spatial frequency features (x-axis).\\
The fill factor ($\zeta$) displayed in \cref{fig:probe profile} was calculated by counting the number of pixels in the matrix with a value different than 0 and dividing it by the total number of pixels a round aperture with the same radius will have. Furthermore, the spacing between holes shown in \cref{fig:phaseplatesketch} is arbitrarily reduced for simplicity.
The calculated probe size ($d_{50}$) for the different proposed designs is shown in \cref{fig:probe size}. We see the simulated $d_{50}$ value for each of the plates with an increasing opening angle. We found that all the proposed designs offer some $C_s$ correction. However, the linear phase profile obtained by applying first-order correction can keep the phase relatively flat for bigger opening angles, further reducing the probe size. More specifically, we reach a $d_{50}$ value of 0.93\AA~ at 15~mrad for the zeroth-order phase plate, a $d_{50}$ of 0.66\AA~ at 21~mrad for the hybrid design, and a $d_{50}$ of 0.75\AA~ at 18~mrad for the simplified hybrid design. These values represent a ~40$\%$, ~57$\%$, and ~52$\%$ improvement in spatial resolution, respectively, compared with a non corrected instrument. At higher aperture values (i.e., larger than ~21~mrad), we must reduce the width of the segment in order to reduce the phase error, and this will eventually become an issue in terms of fabrication.\\
The relative current density can be calculated from eq.\ref{currdens} and is plotted in fig.\ref{fig:current density} assuming $I_{0}=$50~pA, all the proposed designs increase this value. More specifically, 6.4x for the zeroth-order phase plate, 28x for the hybrid phase plate shown, and 16.4x for the simpler hybrid design compared with a non corrected round aperture at 10~mrad. This relative current density is highly important for, e.g., spectroscopic methods where the increased current in a small probe can lead to a vastly improved signal-to-noise ratio on top of the gain in spatial resolution.
\section{Discussion}
\begin{figure}[b!]
\centering
\includegraphics[width=\textwidth]{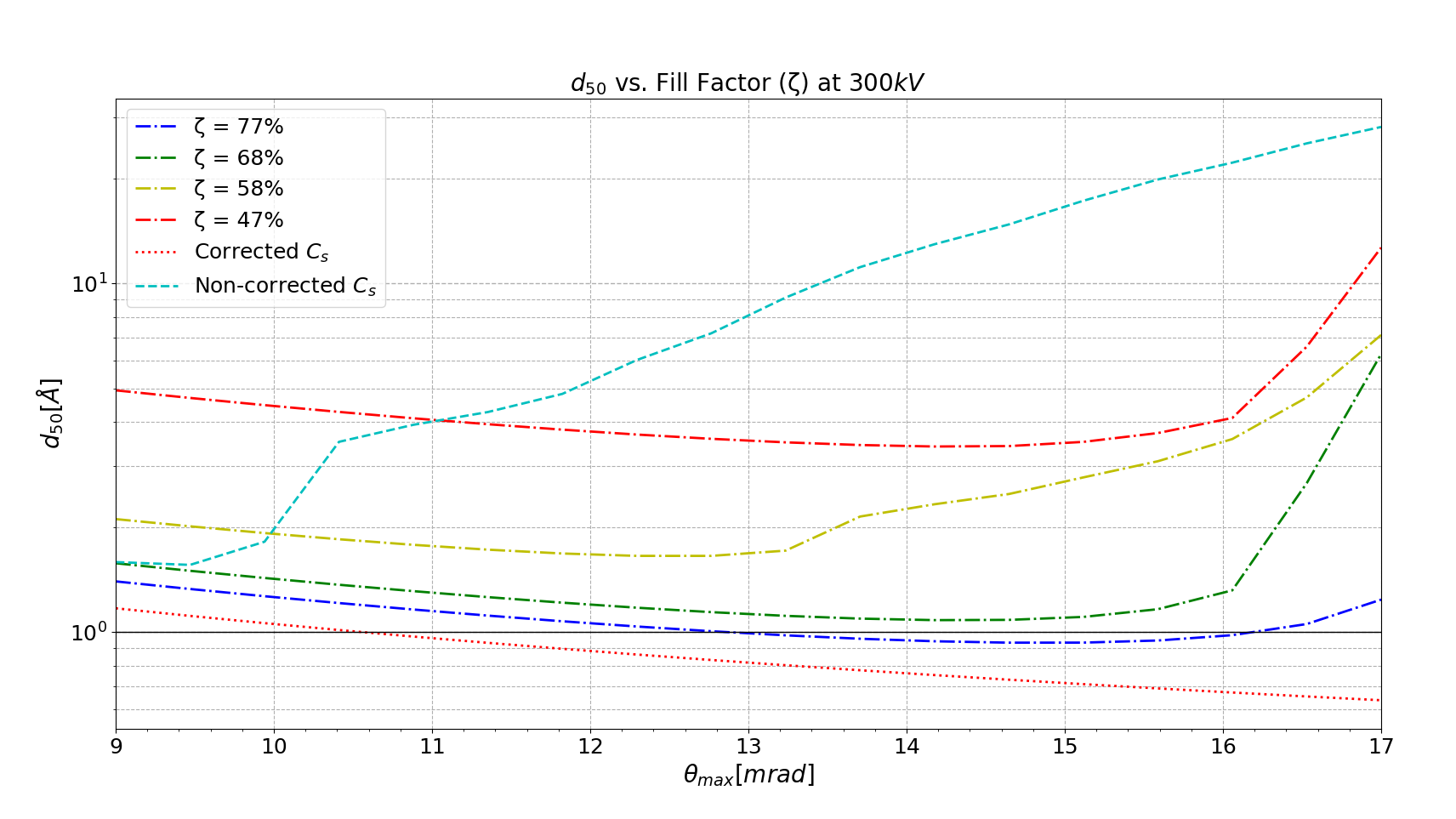}
\caption{\label{fig: probe size vs. fill factor} $d_{50}$ value for different $\zeta$ using a zeroth-order phase plate. We kept the same design as in \cref{fig:probe profile} b. for the segments only increasing the width of the lines going outwards in the radial direction, thus keeping the sampling on $\theta$ the same. We can observe how the $d_{50}$ is heavily affected by $\zeta$, making it difficult of a zeroth-order phase plate to obtain a sub-\AA~ probe for $\zeta<0.70$ }
\end{figure}

\begin{figure}[h!]
\centering
\includegraphics[width=\textwidth]{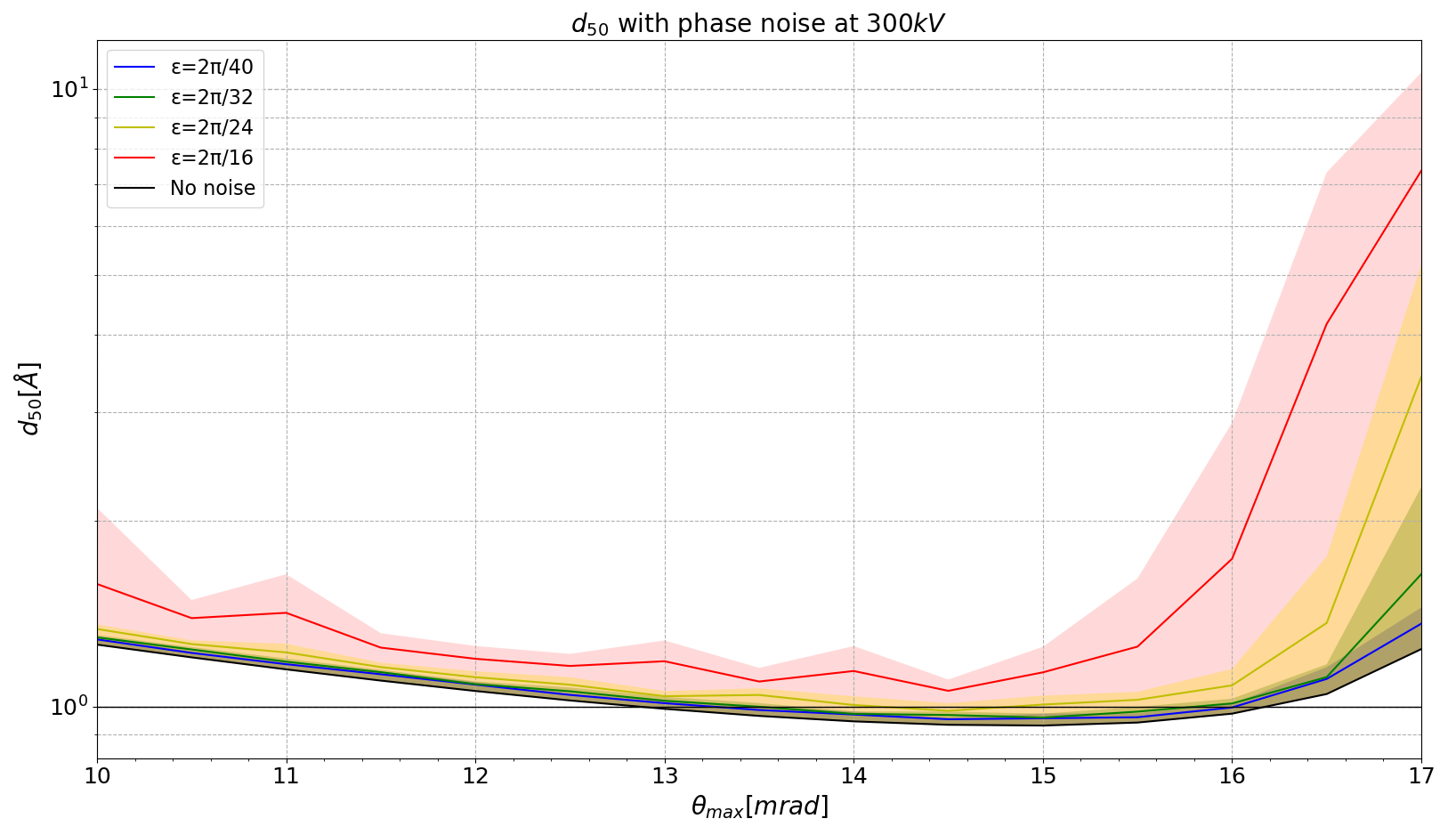}
\caption{\label{fig:Probe Size comparison with noise} Average probe size $d_{50}$ including different levels of phase noise $\epsilon$ for the zeroth-order ring phase plate. The shaded area represents the $2\sigma$ variation for 10 random realizations of the phase noise and the black line shows the performance of the phase plate without any phase noise (lower limit). The plot shows that sub-\AA~ performance is possible for all but the highest noise level simulated here. }
\end{figure}

This design exercise shows that having an adaptive phase plate in the condenser aperture plane can correct $C_{s}$. Not only did we numerically obtain a probe size below the 1\AA~ limit, but we also increased the relative current density more than 20x. However, implementing a device like the one proposed in this study poses several challenges. The most critical issue is the possibility of manufacturing a device with all the necessary electrical connections to control each phase segment separately. As the aperture angle increases, we quickly approach regions where the aberration function changes rapidly; this change requires a narrow segment to keep the phase error within a reasonable range. However, reducing this segment size or going from zeroth-order segments to first-order segments will increase the number of interconnections needed to control such implementation, ultimately reducing the attainable fill factor.\\
We show the relation between probe size ($d_{50}$) and fill factor ($\zeta$) in \cref{fig: probe size vs. fill factor}. We can observe how the probe size for a plate with a $\zeta<0.5$ cannot even match the performance of a non corrected instrument in terms of $d_{50}$ for small angles, but at angles $>11$~mrad this less than ideal phase plate can still significantly improve $d_{50}$ and hence increase $J_{50}$ when compared with the non corrected situation. This may be of interest in cases where current density is more important than ultimate resolution.\\
It is important to note that the fill factor has a double negative effect on the probe current. On the one hand, it lowers the current in the beam due to partial blocking by a factor $\zeta$. On top of this, the beam that gets through is split into the desired part (central spot) and a tail part, which further lowers the intensity of the desired part of the beam to $\zeta^2$. Often there is more than enough beam current available, and the sample may limit how much current can be used. In such cases, the initial loss due to finite transparency of the phase plate is not a problem, but the tail argument still is, as it will provide a degraded image contrast while still doing full probe current beam damage.
Increasing the number of segments (more interconnections) or reducing their size brings the fabrication process closer to the photolithography resolution limit ($\approx 1\mu m$) as indicated in fig.\ref{fig:min. feature size}. For this reason, one must be careful with the phase plate design since high complexity designs will require higher-resolution methods such as EUV or e-beam lithography \cite{engstrom2014additive}. For instance, a zeroth-order correction requires only one electrode to create an Einzel lens inside the region of the segment, while a first-order implementation would require a plate capacitor-like arrangement to achieve the required phase ramp. Another way to increase the fill factor is to tie the different electrodes together with a fixed resistor string, saving multiple interconnection lines. The drawback of this solution is that it would take away the ability to fine-tune each segment individually, and if the projected potentials are not quite what was expected, there is no easy way to correct them unless resistor values could be, e.g., laser trimmed. Furthermore, having a significant $C_s$ correction seems unreasonable for phase plates with less than $50\%$ fill factor since most of the current will land on the tails of the probe, as shown above. However, we demonstrate here that the number of segments needed to achieve sub-1\AA~ correction can be kept relatively low, thus reducing the design complexity significantly over, e.g., the design proposed in our previous paper \cite{verbeeck2018demonstration}. The predicted resolution should be taken with some reservation as effects like source size broadening, vibrations, chromatic aberration, higher-order aberrations, and other non-idealities were not considered for the theoretical study and will further lower the actual attainable resolution in practice.\\
An advantage of the proposed designs is that, as indicated earlier, the performance of an electrostatic phase plate is relatively insensitive to the quality of the voltage sources driving the segments  \cite{verbeeck2018demonstration}. We test this claim in the zeroth-order design in \cref{fig:Probe Size comparison with noise} by adding different normally distributed phase noise to each of the phase segments and calculating the resulting $d_{50}$ and the difference between a case where no noise is present. We note that the increase in probe size to such noise remains under 0.1\AA~ for most aperture angles. Assuming an electronic system is designed to provide, e.g., a maximum of $200\pi$ phase shift, such precision and noise requirements would easily be met even, e.g., with a humble 12-bit digital to analog converter.

\section{Conclusion}
We have numerically demonstrated how a programmable electrostatic phase plate can correct third-order spherical aberration in the TEM with phase plates of modest complexity consisting of 19 segments (zeroth-order) and as few as 8 segments for a hybrid design. All the proposed designs are capable of flattening the electron wavefront up to relatively high opening angles and can provide sub-1\AA~ probe sizes. \\
We discussed the benefit of moving from zeroth-order to first-order phase shifters to contain the phase error within some error margin. On top of this, we showed how the segment width for a first-order element does not necessarily need to have unreasonably small dimensions to correct $C_{s}$ with acceptable performance. In terms of shape, we naturally confirm that a circularly symmetric design compensates for the phase shift caused by $C_{s}$ since it mimics its symmetry; thus, correcting the aberrations more efficiently within each segment. Nevertheless it is likely that in practice also some breaking of this symmetry would be useful to compensate for non round aberration. This would bring a further increase in the number of phase-shifting elements and the complexity of getting interconnects to each.\\
We found that to achieve $C_{s}$ correction to any significant extent, the phase plate has to have a fill factor $\zeta \geq 0.75$ to achieve $d_{50} \leq 1$\AA~.\\
We demonstrated a significant increase in current density, crucial for applications such as STEM EELS, even for cases where the fill factor is low. When the absolute current density is important, we argued that it scales as $\zeta^2$ as both the total current and the amount of probe tails scale with the fill factor.\\
We investigated the robustness of the correction versus noise on the electrostatic potentials of the phase-shifting elements and showed that, for a zeroth-order phase plate, we get excellent results for $\epsilon \leq 2\pi/24$, which is well within reach of even simple digital-analog converter circuits. \\
This study demonstrates how an adaptive optical device can potentially be implemented in a non corrected instrument, improving its performance without any significant changes in the optical setup of the column. Having adaptive optical elements in an electron microscope allows for increased flexibility and performance, which opens up a wide variety of experimental setups, ultimately expanding the microscope's capabilities.\\ 
Besides (adaptive) aberration correction, one of the potential applications for electrostatic phase plates is to shape the beam to either enlarge the depth of focus or increase the z-resolution. This first idea has been demonstrated before using a spatial light modulator that can generate Bessel-like beams invariant with propagation length \cite{chattrapiban2003generation,tao2003dynamic}. However, we can also attempt to correct higher-order aberrations with a phase plate to increase the opening angle, potentially enhancing the z-resolution (which scales as the inverse of the aperture squared).\\
 Another possible application is to use the phase plate for phase retrieval experiments. This idea has already been studied in optics \cite{zhang2007phase,zhang2010phase,falldorf2010design}, and the possibilities of fast and reliable wavefront tuning with the electrostatic phase plate can allow us to do the same in the electron microscope.\\
 Increasing selectivity in inelastic scattering experiments by differential experiments changing the probe rapidly between two or more configurations is another class of applications that could shed light on, e.g., the magnetic, chiral, and optical response of materials at the nanoscale. Furthermore, having such adaptive apertures allows for automatic correction and optimization, self-tuning the phase of each of the segments to match the user's needs.\\
However, it is evident that the performance of the proposed designs in terms of $C_{s}$ correction capabilities is inferior to that of a modern multipole corrector. Still, the proposed setup would offer several significant advantages, such as small size (1 mm scale), low power consumption (1 W scale), high speed (up to 100 kHz, extendable to much higher), no hysteresis, vastly reduced precision constraints on drive electronics (12 bit suffices), negligible drift, and potentially low production cost.\\
These observations will guide further practical implementations with experimental realization of phase plate-based aberration correction on the nearby horizon.

\section*{Acknowledgements}

All authors acknowledge funding from the Flemish Research Fund under contract G042820N "Exploring adaptive optics in transmission electron microscopy". J.V. acknowledges funding from the European Union's Horizon 2020 Research Infrastructure - Integrating Activities for Advanced Communities under grant agreement No 823717 – ESTEEM3 and from the University of Antwerp through a TOP BOF project.

\bibliographystyle{abbrv}
\bibliography{sample}

\end{document}